\documentclass[aps,paper,nofootinbib, superscriptaddress]{revtex4}

\usepackage[english]{babel}
\usepackage[T1]{fontenc}
\usepackage{amsmath}
\usepackage{latexsym}
\usepackage{dsfont}
\usepackage{amsfonts}
\usepackage{graphicx}

\newcommand{\be}{\begin{eqnarray}}
\newcommand{\ee}{\end{eqnarray}}
\newcommand{\bite}{\begin{itemize}}
\newcommand{\eite}{\end{itemize}}
\newcommand{\la}{\langle}
\newcommand{\ra}{\rangle}

\newcommand{\dd}{\mathcal{D}}

\begin{document}
\title{Quantum Interactions Between Non-Perturbative Vacuum Fields}
\author{R. Millo}
\affiliation{Universit\'a degli Studi di Trento and I.N.F.N.\\ Via Sommarive 14, Povo (Trento), Italy.}
\author{P. Faccioli}
\affiliation{Universit\'a degli Studi di Trento and I.N.F.N.\\ Via Sommarive 14, Povo (Trento), Italy.}
\author{L. Scorzato}
\affiliation{European Centre for Theoretical Studies in Nuclear Physics and Related Areas (E.C.T.*)\\Strada delle Tabarelle 286, Villazzano (Trento), Italy}
\begin{abstract}
%A number of lattice studies have suggested that the non-perturbative QCD dynamics may be driven by specific families of vacuum field configurations, including  
%e.g. instantons, monopoles and center-vortices. 
We develop an approach to investigate the non-perturbative dynamics of quantum field theories, in which specific vacuum field fluctuations are treated as the low-energy dynamical degrees of freedom, while all other vacuum field configurations are explicitly integrated out from the path integral.
We show how to compute the effective interaction between the  vacuum field degrees of freedom
both perturbatively (using stochastic perturbation theory) and fully non-perturbatively (using lattice field theory simulations). 
The present approach holds to all orders in the couplings and does not rely on the semi-classical approximation. 
\end{abstract}

\maketitle

%\flushbottom

%\maketitle

\section{Introduction}

Lattice field theory represents the only available  \emph{ab-initio} framework, which allows to compute matrix elements  of a large class of quantum field theories, in a fully non-perturbative way. 
In particular, due to the continuous advance in the development of new machines and new algorithms, lattice calculations for QCD are now beginning to explore the chiral regime and are already  producing accurate  results for a large class of observables. 

On the other hand, lattice simulations do not directly explain 
the qualitative physical mechanisms which are responsible for  the non-perturbative phenomena. 
It is therefore important to continue developing alternative approaches,  
which can provide physical pictures and direct insights into the qualitative mechanisms.

In the specific context of QCD,  a large effort has been made in the last  decades, in order to identify relevant  low-energy vacuum gauge field configurations, which  are responsible for hadron structure, by driving the breaking of chiral symmetry  and producing color confinement. 
For example, instantons have been shown to play an important role in  the  breaking of chiral symmetry~\cite{chiralsymmetrybyinstantons} and instanton models~\cite{shuryakrev} have been successfully used to predict physical properties of light hadrons (see e.g. ~\cite{mass} ,\cite{ew} and references therein). 
Similarly, vacuum fields made from  monopoles~\cite{monopoles}, center-vortices~\cite{vortices}, merons~\cite{merons}, and, recently, regular gauge instantons~\cite{NegLenz} 
have been shown to generate an area law for the Wilson loop, hence to produce color confinement.  

Once a set of important  low-energy vacuum field configurations 
has been identified, it is natural to address the question whether it is possible  to build an   effective theory, based on such degrees of freedom.
In practice, this corresponds to deriving an expression for the original generating functional, in which the functional integral is restricted to the configurations of the selected family of low-energy  vacuum fields,  while all other field configurations are integrated out and give raise to an   effective interaction.  

In the present paper, we take a step in such a direction. The main idea is to use lattice simulations to generate a statistically representative ensemble of field configurations. Such configurations are then projected  onto the functional manifold formed by chosen the family of vacuum field configurations.  This procedure is conceptually analog to the technique adopted in statistical mechanics to evaluate the free energy, as a function of a
 set of (order) parameters.
The result is a new exact expression of the original path integral, given in terms of an integral over the collective coordinates of the low-energy vacuum field manifold. 

In order to introduce the formalism and illustrate how the approach works, in this first work we  consider the simple case of  a one-dimensional quantum mechanical particle, interacting with a double-well potential.
The choice of such a toy-model is motivated by two facts: on the one hand, the relevant non-perturbative vacuum field configurations for this system are well known: they are  the instantons and anti-instantons, which describe the tunneling between the two classical vacua.  On the other hand,  the simplicity of the model allows us to perform detailed numerical simulations and test our method.

The paper is organized as follows. 
In section \ref{EA},  we introduce our framework for a generic quantum mechanical system. From section \ref{doublewell},  we focus on the  specific case of the double well problem. In particular, in sections \ref{pert} and \ref{nonpert}, we  perform  perturbative and non-perturbative calculations of the instanton-antiiinstanton  effective interaction. In \ref{test} we discuss the results of the numerical implementation of  this method.%  is then  Our main results, conclusions and perspectives are reviewed in section \ref{conclusions}.

%In the next sections, we shall see that the same program can be carried out for computing the effective action in quantum systems, by exploiting the standard formal analogy between quantum mechanical path integrals and canonical partition functions.
%We shall first re-write the quantum mechanical path integral in a form analog to  Eq.(\ref{Z1}), where the role of reaction coordinates $\gamma_1,\ldots, \gamma_n$ is played by the curvilinear coordinates, which  parametrize a family of vacuum field fluctuations. For example, in the simple case of a one-dimensional double-well potential, these can be assumed to be made by superimposing 
Then, we shall use path integral Monte Carlo simulations to generate an un-biased  ensemble of equilibrium field configurations and develop a technique to project such configurations onto the vacuum field  manifold. It is important to stress the fact that this method does not rely on saddle-point arguments.
%the choice of "reaction coordinates" is quite arbitrary and that we  do not make use of  saddle-point analysis. 
%The choice of instantons as low-energy fields  for the double well problem is dictated by the specific physics of the problem, but  it is not in principle mandatory for the present formalism.

\section{Effective Interaction for the Vacuum Field Configurations}
\label{EA}

%In this section, we introduce our projection formalism, which allows  to compute the effective interaction for the low-energy degrees of freedom, which describe the configurations
%of the relevant  vacuum field fluctuations. 
For sake of simplicity, in this work we shall introduce our formalism for a system consisting  of a quantum mechanical particle, interacting with an external potential. 
However, the same  method can be applied to quantum field theories with arbitrary number of dimensions, as long as they can  be formulated on the lattice.  

After performing the Wick rotation to imaginary time, the path integral for the system described by the interaction $U(x)$
and  corresponding to the boundary conditions 
\be
x[-T/2]=-x_i \qquad x[T/2]=x_f
\label{BC1}
\ee
is given by 
\be
Z[x_f,x_i| T] = \langle x_f | e^{- H T} | x_i \rangle =  \int_{x[-T/2]=x_i}^{x[T/2]=x_f}~\mathcal{D} x~ e^{-\frac{S[x]}{\hbar}},
\label{PI}
\ee
where
\be
\label{SE}
S[x]=\int_{-T/2}^{T/2}~\mbox{d}t\,\left[m \frac{\dot{x}^2(t)}{2}+U(x)\right]
\ee
is the usual Euclidean action.

Let us  consider a generic family of vacuum 
field configurations (i.e. of paths) $\tilde x (t;\gamma)$ , which depend on a {\it finite} set of parameters $\gamma=(\gamma_1,\ldots,\gamma_k)$ and satisfy the boundary conditions
(\ref{BC1}).
The paths $\tilde x (t;\gamma)$ form a differentiable manifold $\mathcal{M}$, parametrized by the curvilinear coordinates $\gamma_1, \ldots, \gamma_k$. 

For every given choice of the  parameters $\gamma$ it is possible to  decompose a generic path $x(t)$ contributing to the path integral (\ref{PI}) as a sum of a
field configuration $\tilde x(t;\gamma)$, belonging to the  manifold $\mathcal{M}$, and of a residual field  $y(t)$: 
\be
\label{decompose}
x(t) \equiv \tilde x(t;\gamma) + y(t).
\ee
We shall refer to the field $y(t)$ as to the "fluctuation field". However, in the following we shall never require that the vacuum field $\tilde x(t;\gamma) $ satisfies the Euclidean classical Eq. of motion (EoM).  
Hence, both $\tilde x(t;\gamma) $ and $y(t)$ represent in general quantum vacuum fluctuations.

Let us now derive a particular representation of the path integral (\ref{PI}) in terms of a set of ordinary integrals over the parameters $ \gamma_1,...,\gamma_k$ and a functional integral over the fluctuation field, $y(\tau)$. 
Since the new representation of the path integral contains $k$ additional integrals over $d\gamma_1,\ldots, d\gamma_k$,  we need to impose $k$ constraints. 
We choose to enforce the $k$ orthogonality conditions
\be
\label{ort}
\left(y(t) \cdot g_{\gamma_i} (t,\bar\gamma) \right) \equiv \int_{-T/2}^{T/2} dt~ y(t) ~ g_{\gamma_i} (t,\bar\gamma)=0, \qquad i=1,\ldots, k
\ee
where the functions $g_{\bar \gamma}^{i}(t)$ are defined as
\be
g_{\gamma_i} (t,\bar\gamma) = \left.\frac{\partial}{\partial \gamma_i} \tilde x(t;\gamma)\right|_{\gamma=\bar \gamma}.
\label{ggamma}
\ee

In order to clarify the meaning of the condition (\ref{ort}) we observe that the functions $\{g_{\gamma_i} (t,\bar\gamma)\}_{i=1,\ldots,k}$ identify the $k$ directions tangent to the manifold $\mathcal{M}$ of vacuum fields, 
in the point of curvilinear coordinates  $\bar \gamma=(\bar \gamma_1,\ldots,\bar\gamma_k)$.
 %{\bf Notice that, in general, there may points of the manifold in which the tangent space is not defined. We will discuss this situation in the next sections.}   
%Note that $y(t)$ and $\gamma$ in Eq. (\ref{decompose}) depend implicitly on $\bar \gamma$.  
We consider only choices of manifold and $\bar \gamma$ such that the vectors (\ref{ggamma}) define a system of coordinates on the manifold.
The coordinates $(\Psi_1,\ldots, \Psi_k)$ of a point $\tilde x(t;\gamma)$    are defined as:
\be
\Psi_1[\tilde x(t;\gamma)] & = & (\tilde x(t;\gamma)\cdot g_{\gamma_1} (t,\bar\gamma) ) 
\label{lp1}\\
 &...&\nonumber\\
\Psi_k[\tilde x(t;\gamma)] & = & (\tilde x(t;\gamma)\cdot g_{\gamma_k} (t,\bar\gamma) ).
\label{lpk}
\ee
Configurations which lie in a functional  neighborhood  of the manifold can be projected onto the same system of coordinates.  The components of such paths $x(t)$ are
\be
\Psi_1[x(t)] &=&( x(t)\cdot g_{\gamma_1} (t,\bar\gamma) ) =  \int_{-T/2}^{T/2} dt~ x_i(t)~g_{\gamma_1} (t,\bar\gamma)\label{j1}\\
 &...&\nonumber\\
\Psi_k[x(t)] &=& ( x (t)\cdot g_{\gamma_k} (t,\bar\gamma)) = \int_{-T/2}^{T/2} dt ~x_i(t) ~g_{\gamma_k} (t,\bar\gamma).
\label{jk}
\ee
Hence, the condition (\ref{ort}) imposes that fluctuation fields $y(t)$ should have vanishing coordinates on the system of coordinates defined by the vector $\{g_{\gamma_i} (t,\bar\gamma)\}_{i=1,\ldots,k}.$

% It is usually convenient to choose the manifold point $\gamma  = \bar\gamma$ 
%to be a solution of the classical EoM. In fact--- as we shall see in section \ref{pert}---with  this choice, the effective action for $\gamma\simeq\bar\gamma $ can be evaluated perturbatively.
 %As we shall see, with  such a choice, it is possible to develop small fluctuations around it.

Let us now apply a standard technique to implement the $k$ constraints (\ref{ort}) inside the path integral (\ref{PI})~ \cite{diakonov, hutter}.  
We introduce a Faddeev-Popov unity:
\be
\label{fadeev}
1= \int d^k \gamma \int \dd y ~\left(\prod_i \delta^{(k)}(y(t) \cdot g_{\gamma_i} (t,\bar\gamma) \right) ~\delta[ \tilde x(t;\gamma) + y(t) - x(t)] ~\Phi[x],
\ee
which serves as a definition of the functional $\Phi[x]$. 
Note that the integration on $y(t)$ in (\ref{fadeev}) can be trivially performed and one obtains
\be
\label{phim1}
\Phi^{-1}[x] = \int d^{k} \gamma'~ \prod_i \delta^{(k)}\bigg( (x(t)-\tilde x(t;\gamma')) \cdot g_{\gamma_i} (t,\bar\gamma) \bigg)
\ee
In particular, we are interested in the value of $\Phi[x]$ at the point $x(t)= \tilde x(t;\gamma) + y(t)$. If we insert (\ref{fadeev}) in the original path integral (\ref{PI}), we obtain, after integration over $x$: 
\be
\label{PI2}
Z(T;x_i,x_f) = \int \prod_{l=1}^k~ d \gamma_l \int_{y(-T/2)= 0}^{y(T/2)= 0} \dd y ~
~\left(\prod_i \delta^{(k)}(y(t) \cdot g_{\gamma_i} (t,\bar\gamma) \right)~\Phi[\tilde x(t;\gamma)+y(t)]
%\left| \det_{ij} \left(  g^i_{\bar\gamma}(t) \cdot \frac{\partial \tilde x(t, \gamma)}{\partial \gamma^j}\right)\right|
e^{-\frac{S[\tilde x(t;\gamma)+y(t)]}{\hbar}},
\ee
where the dependence the initial and final points $x_i$ and $x_f$ enters implicitly, through the vacuum field $\tilde x(t;\gamma)$.

\begin{figure}[t]
\begin{center}
\includegraphics[width=9cm]{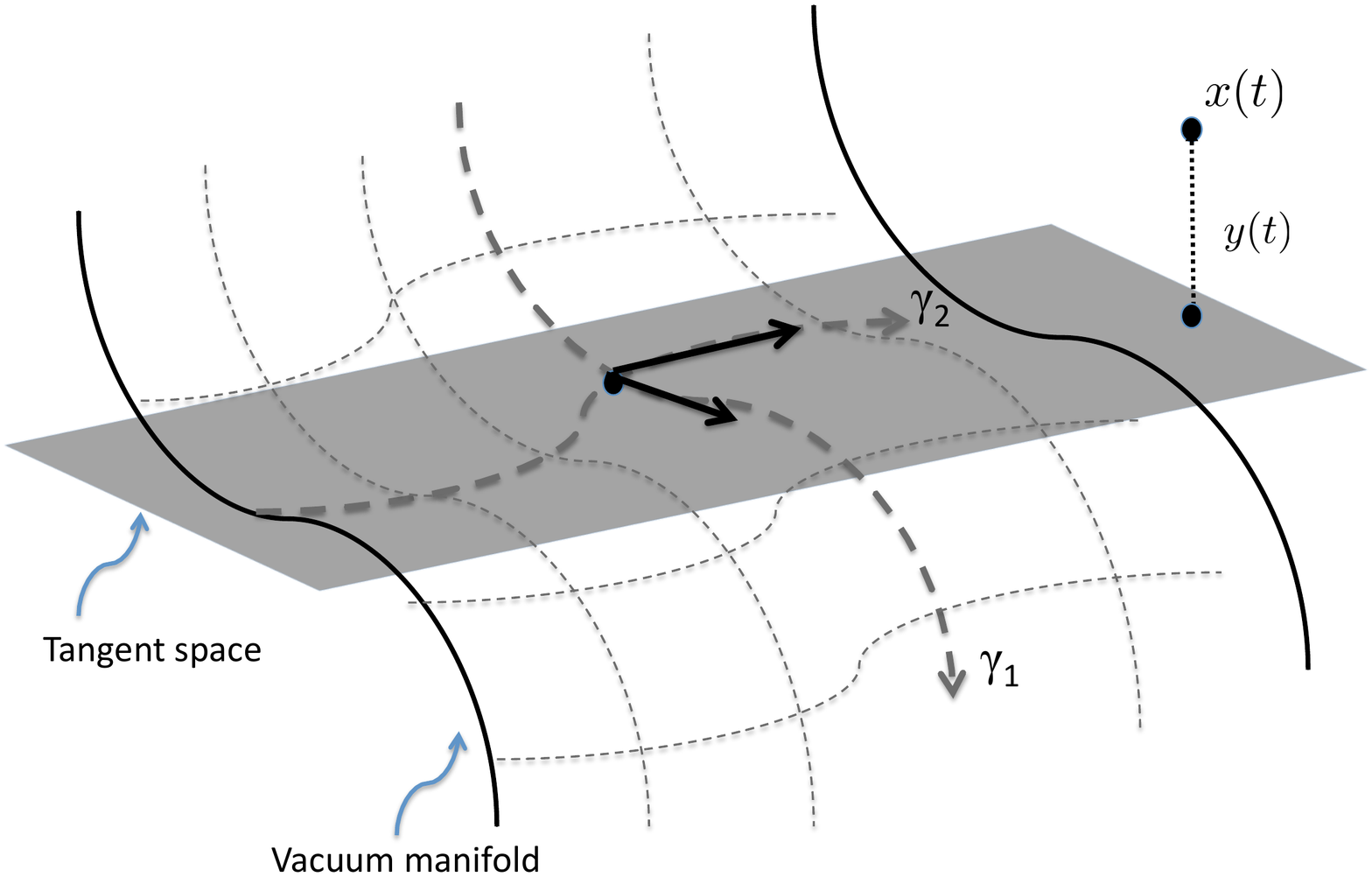}
\caption{Pictorical representation of the projection of the path $x(t)$ to onto the vacuum field manifold.  A path $x(t)$ is represented by a point in this picture. The constraints
(\ref{ort}) imply that the
the fluctuation field $y(t)$ is perpendicular to the plane tangent to the manifold in the point of  the curvilinear abscissas $\gamma=\bar \gamma$. }
\label{geom}
\end{center}
\end{figure}

%If $U(x)$ is a double-well potential, it is natural to choose $\tilde x(t, \gamma)$ as a multi-instanton background. In this case, the parameters $\gamma_1,\ldots,\gamma_k$ corresponds to the positions  in time of each pseudo-particle. However, it should be stressed that the decomposition (\ref{PI2}) is exact, does rely on 
%semiclassical  arguments and does not apply only to instanton solutions, but It holds generally, for arbitrary background fields. 
The path integral (\ref{PI2}) can be formally re-written as 
\be
\label{Gdef}
Z(T;x_i,x_f) = \int \prod_{l=1}^k~d~\gamma_l~ e^{-\frac{1}{\hbar} F(\gamma_1,\ldots,\gamma_k)},
\ee
where $F(\gamma) $ is  defined as 
 \be
\label{Gdef2}
F(\gamma)=  -\hbar \log
\int_{y(-T/2)= 0}^{y(T/2)= 0} \dd y ~~\left(\prod_i \delta^{(k)}(y(t) \cdot g_{\gamma_i} (t,\bar\gamma)) \right)~
%\left| \det_{ij} \left(  g^i_{\bar\gamma}(t) \cdot \frac{\partial \tilde x(t, \gamma)}{\partial \gamma^j}\right)\right|
\Phi[\tilde x(t;\gamma)+y(t)] e^{-\frac{1}{\hbar} S[\tilde x(t;\gamma)+y(t)]}.
\ee

Some comments on what we have done so far are in order.
First of all we stress that  $Z(T;x_i,x_f)$ can be interpreted as the partition function of a system with a finite number of degrees of freedom $\gamma_1, \ldots, \gamma_k$. The term 
$F(\gamma_1,\ldots, \gamma_k)$ 
% Note that, by construction,  such an interaction is independent on the choice of the parameter $\bar \gamma$.
is the analog of the (free) energy in statistical physics an will be referred to as the \emph{effective interaction}. 
%Such an analogy is elaborated further in appendix \ref{inspire}. In particular,  Eq.s (\ref{Gdef}) and (\ref{Gdef2}) are the quantum-mechanical analog of Eq.s (\ref{Z1}) and (\ref{Z2}).

Let us now address the problem of how to compute $F(\gamma_1, \ldots, \gamma_k)$, using lattice simulations. %To this end, we need to implement step 3 through 5 of the algorithm discussed in the previous section.
Let $\{x_1(t), \ldots, x_{N_{conf}}(t)\}$ be a statistically representative ensemble  of $N_{conf}$ paths (i.e. obtained by means of lattice simulations).
 %and let us project each of them onto the tangent space, spanned by the functions $g_{\gamma_1},\ldots,g_{\gamma_k}$:
%\be
%\Phi_1[x_i(t)] &=& ( x_i(t)\cdot g_{\gamma_1} (t,\bar\gamma) ) = \int_{-T/2}^{T/2} dt~ x_i(t)~g_{\gamma_1} (t,\bar\gamma) \label{j1}\\
% &...&\nonumber\\
%\Phi_k[x_i(t)] &=& ( x_i(t)\cdot g_{\gamma_k} (t,\bar\gamma)) = \int_{-T/2}^{T/2} dt ~x_i(t) ~g_{\gamma_k} (t,\bar\gamma),\qquad i=1,\ldots, N_{conf}.\
%\label{jk} 
%\ee
The coordinates $(\Phi_1, \ldots, \Phi_k)$ of each of such paths are specified by the Eq.s (\ref{j1})-(\ref{jk}). 
Using  the definition  (\ref{decompose}) and the orthogonality conditions (\ref{ort}) we  obtain a set of $k$ non-linear Eq.s for the $\gamma_1,\ldots, \gamma_k$ variables:
\be
\Phi_1[x(t)] = (x(t) \cdot g_{\gamma_1} (t,\bar\gamma) )&=&  (\tilde x(t;\gamma)\cdot g_{\gamma_1} (t,\bar\gamma) ) \equiv \Psi_1(\gamma)
\label{pp1}\\
&...&\nonumber\\
\Phi_k[x(t)] = (x(t) \cdot g_{\gamma_k} (t,\bar\gamma) ) &=& (\tilde x(t;\gamma)\cdot g_{\gamma_k} (t,\bar\gamma) ) \equiv \Psi_k(\gamma).
\label{ppk}
\ee
Note that, while the coordinates $\Phi_1, \ldots, \Phi_k$ on the left-hand-side depend on the path $x(t)$,  the functions $\Psi_1(\gamma), \ldots, \Psi_k(\gamma)$ on the right-hand-side  depend only on the set of collective coordinates
 $(\gamma_1, \ldots, \gamma_k)$ and are determined by the choice of the background field manifold and of the parameter $\bar\gamma$. 
Hence, by solving numerically such a system of Eq.s,  a value for the curvilinear coordinates  can be assigned to each configuration. 
Repeating this procedure for the entire ensemble of lattice configurations $x_1(t), \ldots, x_{N_{conf}}(t)$ one  determines the probability density 
$\mathcal{P}(\gamma_1, \ldots, \gamma_k)$,  which relates directly to the effective interaction
\be
F(\gamma_1,\ldots,\gamma_k) = -\frac{1}{\hbar}~\log \mathcal{P}(\gamma_1,\ldots, \gamma_k).
\ee

The effective theory defined by the partition function (\ref{Gdef}) allows to perform approximate calculations of the vacuum expectation value of arbitrary operators $\hat O(t)$,
\be
\langle \hat O(t)\rangle = \frac{\int \mathcal{D} x~ O[x(t)] ~e^{-\frac{1}{\hbar}S[x]}}{\int \mathcal{D} x e^{-\frac{1}{\hbar}S[x]}}.
\label{Oexp}
\ee
In fact, if the vacuum manifold contains the physically important vacuum configurations, then $O[x(t)]\simeq O[\tilde x(t;\gamma)]$ and
\be
\langle \hat O(t)\rangle \simeq \frac{\int \prod_{l=1}^k~d~\gamma_l~ O[\tilde x(\gamma_1, \ldots, \gamma_k)]~
e^{-\frac{1}{\hbar} F(\gamma_1,\ldots,\gamma_k)}}{\int \prod_{l=1}^k~d~\gamma_l~ e^{-\frac{1}{\hbar} F(\gamma_1,\ldots,\gamma_k)}}.
\ee

We note that, while the partition function (\ref{Gdef}) is independent  on the choice of  $\bar \gamma$ --- which specifies the system of coordinates on the manifold---
the effective interaction $F(\gamma)$ and vacuum expectation values of operators may in principle depend on such a parameter. 
However, such a dependence is generated only by the projection of paths which contain very large fluctuations, i.e.  lie far from the vacuum manifold. 
To see this, let us consider the projection 
of a configuration $x(t)$ which lies very close to a point  on the vacuum manifold  $\tilde x(t; \gamma')$, i.e. 
\be
||x(t)-\tilde x(t; \gamma')|| \simeq 0, 
\ee
for some $\gamma'$.
 Then, the projection Eq.s (\ref{pp1})-(\ref{ppk}) read:
\be
 (x(t)\cdot g_{\gamma_1} (t,\bar\gamma) )\simeq  (\tilde x(t;\gamma')\cdot g_{\gamma_1} (t,\bar\gamma) ) &=&  (\tilde x(t;\gamma)\cdot g_{\gamma_1} (t,\bar\gamma) )
\label{pe1}\\
 &...&\nonumber\\
 (x(t)\cdot g_{\gamma_k} (t,\bar\gamma) )\simeq (\tilde x(t;\gamma')\cdot g_{\gamma_k} (t,\bar\gamma) ) &=&(\tilde x(t;\gamma)\cdot g_{\gamma_k} (t,\bar\gamma) ).
\ee 
A solution of such set of Eq.s is trivially $\gamma'=\gamma$, for any choice of the projection point $\bar \gamma$.
If the vacuum field manifold captures the physically important  configurations,  the vacuum expectation values of operators $\hat O(t)$ will be dominated by the
configurations in the functional vicinity of the manifold. 
In the limit in which the relevant configurations are only those very close to the manifold, the system of Eq.s (\ref{pp1})-(\ref{ppk}) have a unique solution and the expressions (\ref{Oexp})
 become independent on the  choice of  the coordinate system on the manifold, i.e. of the parameter $\bar\gamma$.
Clearly, this condition can be verified by comparing the results obtained projecting onto different points of the manifold.

In the rest of this paper, we shall illustrate how this method is implemented in practice, in the specific case of the one-dimensional quantum double-well problem.

\section{Application to the quantum mechanical double well problem}
\label{doublewell}

The discussion made so far has been  completely general:  Eq.s (\ref{Gdef}) and (\ref{Gdef2}) hold for an arbitrary choice of the potential $U(x)$, of the  vacuum field manifold 
$\mathcal{M}$ and
of the boundary conditions $x_i$ and $x_f$. As an illustrative example, let us now restrict our attention to the specific system defined by the potential
\be
\label{U}
U(x)= m \alpha \left( x^2 - \beta^2 \right)^2,
\ee
where $m$ is the mass of the particle. We consider the path integral with periodic boundary conditions (see Eq.~ \ref{BC1})
\be
\label{BC3}
x_i=x_f= - \beta~\qquad\textrm{( or equivalently:}~x_i=x_f= + \beta~).
\ee
In this specific system, a choice of the effective degrees of freedom is suggested semi-classical arguments. We choose the vacuum field manifold to be the one generated by the  
  superposition of $N$ instantons and $N$ antiinstantons. Obviously, the choice of the optimal number of pseudo-particles depends on the time interval $T$.  If the barrier is sufficiently high, one can fix $N$ from the semi-classical\footnote{In QCD, where a strict semi-classical analysis cannot be  consistently applied, the number of pseudo-particles may be estimated form phenomenology or lattice simulations.} tunneling rate~\cite{colman}
\be
2 N \simeq \kappa  T,
\ee  
where $\kappa$ is the one-instanton measure
\be
\kappa \simeq 4 \sqrt{\frac{2 (2 \alpha)^{3/2} \beta^2}{\pi \hbar}}.
\ee

The curvilinear coordinates $\gamma_1=t_1,\gamma_2=\bar t_1, \ldots, \gamma_{2N-1}=t_N, \gamma_{2N}=\bar t_N$ represent the collective coordinates of each instanton or antiinstanton, i.e. their positions in the imaginary time axis.  
In particular, we adopt the so-called "sum-\emph{ansatz}", which consists in simply adding-up  the instanton and antiinstanton fields:
\be
\label{sum4}
\tilde{x}_{S_{2N}}( t ; t_1, \ldots \bar t_N ) \equiv -\beta+\sum_{k=1}^N\left[\hat{x}_{I}(t-t_k)+\hat{x}_{\bar{I}}(t-\bar t_k)\right],
\ee
where we have labeled with  $t_1,\ldots t_N$~($\bar t_1,\ldots, \bar t_N$) the centers of the instantons (antiinstantons). 
The path integral, re-written as in Eq.(\ref{Gdef}) reads
\be
\label{ZILM}
Z[T;-\beta, -\beta]= \int dt_1, \int d\bar t_1 \ldots \int d t_N\int d \bar t_{N} ~e^{-\frac{1}{\hbar}~ F(t_1,\bar t_1,\ldots,t_N, \bar t_{N})}
\ee

We note that there are only two choices of such  collective coordinates for which the field configuration (\ref{sum4}) becomes an exact solution of the Euclidean EoM:
\begin{enumerate}
\item When all nearest neighbour instanton-antiinstantons pairs are infinitely separated from each other, i.e. 
$$|t_i-\bar t_i|\to \infty, |t_{i+1}-\bar t_i| \to \infty $$
\item When all nearest neighbour instanton-antiinstantons pairs are infinitely close to each other, i.e. 
$$t_i=\bar t_i$$.
\end{enumerate}
In the former case, one obtains a dilute instanton gas configuration. In the latter case, all pairs annihilate and the field reduces a trivial classical vacuum, i.e.
$x(t)=-\beta$. For any other choice of the collective coordinates $t_1, \ldots, \bar t_N$, the  field configuration (\ref{sum4}) is   not an extremum of the action.

The relative statistical weight of each configuration in the path integral (\ref{PI}) is provided by the exponential factor appearing in Eq.~(\ref{ZILM}), which plays the role of the free energy in the statistical mechanical analogy.
Hence, the function $F( t_1,\ldots, \bar t_N)$ expresses the statistical and dynamical correlations between the pseudo-particles, induced by all other field configurations in the path integral. For example, in the high barrier limit in which the semi-classical dilute instanton gas approximation is justified, one has
\be
e^{-\frac 1\hbar~F(t_1, \ldots, \bar t_N)} \simeq \theta(\bar t_1-t_1)~\theta(t_2-\bar t_1)\ldots \theta(\bar t_N - t_{N})~\kappa^{2 N}. 
\label{FDIGA}
\ee

As the height of the barrier is adiabatically reduced, the dilute instanton gas approximation becomes worse and worse and eventually  breaks down. In this regime,
  the vacuum fields behave as an interacting liquid and the effective interaction $F(t_1, \ldots, \bar t_N)$ deviates from the 
 expression (\ref{FDIGA}) and can be written as 
\be
\label{2body}
F(\bar t_1,\ldots, \bar t_N) \simeq  \sum_{i=1}^{N} F^{IA}_{2}(\bar t_{i}-t_i) + F^{AI}_{2}(t_{i+1}-\bar t_{i}),
\ee
where $F^{IA}_2$ ($F^{AI}_2$) expresses the  two-body instanton-antiinstanton (antiinstanton- instanton) correlations~\footnote{ Eq. (\ref{2body}) can be generalized to include higher-order 
(e.g. three-body, four-body, etc...) correlations.}.
 For very low barriers,  the average instanton distance becomes smaller than the instanton size, and the pseudoparticles "melt". Clearly, in such a regime, instantons and antiinstanton fields no longer represent a good choice of low-energy vacuum degrees of freedom. In the remaining of this work, we shall consider systems for which the dilute liquid regime is appropriate.

In order to compute $F^{IA}_2(t'-t)$ and $F^{AI}_2(t'-t)$  it is convenient to integrate out from (\ref{ZILM}) all instanton degrees of freedom, except those of a single pair
of pseudo particles. To this end, we rewrite the path integral as :
\be
Z[T;-\beta, -\beta] &=& \frac{1}{2}\left[\int d t_1 \! \int d \bar t_1\left( \int d t_2 ~ d\bar t_2 \ldots d\bar t_N~e^{-\frac{1}{\hbar}F}\right) + 
\int d \bar t_1 \int d  t_2~\left( \int d t_1 \,d\bar t_2\,  d t_3 \ldots d\bar t_N~e^{-\frac{1}{\hbar}F}\right) \right]\\
&=&\frac{1}{2}~\left( \int d t'\int d t~ g_2^{IA}(t'-t) + \int d t  \int d t' ~g_2^{AI}(t' -t)\right).\label{translinvar}
% \int_{-T/2}^{T/2} d t_1 \int_{-T/2}^{T/2} d t_2~ g_2^{IA}(t_2-t_1) + \int_{-T/2}^{T/2} d t_1 \int_{-T/2}^{T/2} d t_2 ~g_2^{AI}(t_2-t_1).
\ee
 The first term  corresponds to the case in which the pseudo-particle of coordinate $t$ is an instanton, while the second of coordinate $t'$ is an anti-instanton
  and $g_2^{IA}(t'-t)$ is the corresponding pair-correlation function. Conversely, the second term corresponds to the case in which the  pseudo-particle at $t$ 
  is an anti-instanton and that at $t'$ is an instanton.  
In the dilute liquid regime, the functions  $g^{IA(AI)}(t'-t)$ relate directly to $F^{IA(AI)}_2(t'-t)$ by 
\be
e^{-\frac 1 \hbar ~F^{IA(AI)}_2(t'-t)} \propto~g_2^{IA(AI)}(t'-t),
\ee
where the proportionality factor is controlled by the density.

%\section{Projection onto the Vacuum-FIeld Manifold}
%\label{project}

In order to extract the instanton-antiinstanton pair correlation function $g^{IA}_{2}$ we consider the path integral with boundary condition $x_f=x_i=-\beta$ and parametrize a generic configuration $x(t)$ using the sum ansatz for an instanton-antiiinstanton pair, Eq.(\ref{sum4})
\be
x(t)&=&\tilde{x}^{IA}_{S_2}\left(t; t_1,t_2 \right)+y(t)\\
&=&-\beta\left\{1-\tanh\!\left[\sqrt{2\alpha}\,\beta\,(t-t_1)\right]+\tanh\!\left[\sqrt{2\alpha}\,\beta\,(t-t_2)\right]\right\}+y(t)
\ee
where $y(t)$ is a configuration of boundary conditions $y(\pm T/2)=0$, and $t_1$ and $t_2$ are the coordinates of the two pseudoparticles, in the Euclidean time axis. 
Conversely, in order to evaluate $g_{2}^{AI}$, one should consider the path integral with boundary conditions $x_f=x_i=\beta$ and adopt a vacuum manifold  based on the 
anti-instanton instanton pair:
\be
x(t)&=&\tilde{x}^{AI}_{S_2}\left(t; t_1,t_2 \right)+y(t)\\
&=& \beta\left\{1-\tanh\!\left[\sqrt{2\alpha}\,\beta\,(t-t_1)\right]+\tanh\!\left[\sqrt{2\alpha}\,\beta\,(t-t_2)\right]\right\}+y(t).
\ee
Since the two calculations are  identical, in the following we shall focus on determining $g_{2}^{IA}$ and the $IA$ suffix will be implicitly assumed.

It is convenient to introduce the relative variables 
\be
 \chi &=&\frac{1}{2}~(t_1+t_2),\nonumber\\
 \xi &=&t_2-t_1.\nonumber
\ee
Notice that  variable $\chi$ is the "center of mass" of  the pair, while $\xi$ represents the "relative distance" between the instanton and antiinstanton.  Notice also that Eq.(\ref{translinvar}) implies
\be
F_2(t_1,t_2)=F_2(t_2-t_1)\equiv F_2(\xi),
\ee
that is to say we expect the effective interaction to be independent from the center of mass of the pair. This is a consequence of the time translational invariance of the vacuum.
%Clearly,the time-translational invariance of the vacuum 
%implies that the two-body part of the quantum effective potential must depends only on the relative distance $\xi$: $F_2 = F_2(\xi)$.

%{\bf Da togliere da qui....The path integral reads
%\be
%Z[T;-\beta, -\beta ] &=& \int d t_1~\int d t_2\, \int~\mathcal{D} y~\exp\left\{-\int~\mbox{d}t~\mathcal{L}[\tilde{x}_{S_2}(t; t_1,t_2)+y(t)]\right\}\times\nonumber\\
%&=& \int d \chi~\int d \xi ~\int~\mathcal{D} y~\exp\left\{-\int~\mbox{d}t~\mathcal{L}[\tilde{x}_{S2}(t^{''},-\xi,\xi)+y(t+T)]\right\}
%\label{cambiot}\\
%&& \times~\delta\bigg( y(t)\cdot g_{t_1}\big(t;\bar{t}_1,\bar{t}_2\big)\bigg)~\delta\bigg( y(t)\cdot g_{t_1}\big(t;\bar{t}_1,\bar{t}_2\big)\bigg) \Phi
%T ~\int\mbox{d}\xi~\int~\mathcal{D} \tilde{y}~ \exp\left\{-\int~\mbox{d}t~\mathcal{L}[\tilde{x}_{S_2}(t; -\xi/2,\xi/2)+\tilde{y}(t)]\right\}
%\label{cambioy}
%\ee
%fino a qui.]}
%where the equivalence holds in the infinitely long time limit $T\to \infty$, which we shall always assume to take, at the end of the calculation, and $\tilde{y}(t)=y(t+\chi)$.
%We note that the integration over the $\chi$ variable yields only an overall factor of the total time interval  $T$, which plays the role of the space-time volume in quantum field theory.
%where in Eq.(\ref{cambiot}) we performed the change of variable $t^{'}\rightarrow t^{''}=t^{'}-T$ whose jacobian is equal to $\frac{1}{2}$, and in Eq.(\ref{cambioy}) we perforformed the change of variable $y(t)\rightarrow \tilde{y}(t)=y(t+T)$ whose jacobian is equal to $1$. 

We recall that the multi-instanton field configuration and the fluctuation field have to fulfill the  orthogonality conditions (\ref{ort}), which is enforced in a specific point $\gamma=\bar \gamma$ of the manifold. 
The basis vector of the tangent space of the manifold defined by the sum ansatz (\ref{sum4}) are, for an arbitrary point $\gamma=(\bar t_1,\bar t_2)$
\be
g_{t_1} (t; \bar t_1, \bar t_2) &=& \left.\phantom{\int} \partial_{t_1}\tilde{x}_{S_2}(t; t_1,t_2)~\right|_{{t_1=\bar t_1, t_2=\bar t_2}} = -\sqrt{2 \alpha} \beta^2 \textrm{sech}^2\left[\sqrt{2 \alpha} \beta (t-\bar t_1)~\right]\\
g_{t_2} (t; \bar t_1, \bar t_2) &=& \left.\phantom{\int}\partial_{t_2}\tilde{x}_{S_2}(t; t_1,t_2)~\right|_{{t_1=\bar t_1, t_2=\bar t_2}}= \sqrt{2 \alpha} \beta^2 \textrm{sech}^2\left[\sqrt{2 \alpha} \beta (t-\bar t_2)~\right]
\ee
%The  Eq.s (\ref{ort}), imposed in the point $t_1=0, t_2=0$ read
%\be
%0&=&\left. \int~\mbox{d}t~y(t) ~g_{t_1}(t; t_1,t_2)~\right|_{{t_1=t_2=0}} =-\sqrt{2\alpha}\beta^2\int dt ~y(t)\textrm{sech}^{-2}\!\left[\sqrt{2\alpha}\beta \,t\right],\\
%0&=&\left. \int~\mbox{d}t~y(t) ~g_{t_2}(t; t_1,t_2)~\right|_{{t_1=t_2=0}}  =\sqrt{2\alpha}\beta^2\int dt~y(t)\textrm{sech}^{-2}\!\left[\sqrt{2\alpha}\beta \,t\right].
%\ee
Equivalently, in terms of the  $\chi$ and $\xi$ coordinates, the basis vectors of the tangent space in the generic point  $\gamma=(\xi, \chi)$ read
\be
g_{\chi} (t; \bar\chi, \bar\xi) &=&\left.\phantom{\int} \partial_{\chi}~\tilde{x}_{S_2}\left(t;  \chi - \frac 12 \xi, \chi + \frac 12 \xi \right)~\right|_{{\chi=\bar \chi,\xi=\bar \xi}}\nonumber\\
&=&\sqrt{2 \alpha} \beta^2 \left\{\textrm{sech}^2\left[\sqrt{2 \alpha} \beta
\left(t-\bar\chi-\frac{\bar\xi}{2}\right)\right] - \textrm{sech}^2\left[\sqrt{2 \alpha} \beta \left(t-\bar \chi+\frac{\bar\xi}{2}\right)\right]\right\}\\
g_{\xi} (t; \bar\chi, \bar\xi)  &=& \left.\phantom{\int} \partial_{\xi}~\tilde{x}_{S_2}\left(t;  \chi - \frac 12 \xi, \chi + \frac 12 \xi \right)~\right|_{{\chi=\bar \chi,\xi=\bar \xi}}\nonumber\\
&=&\sqrt{\frac \alpha 2} \beta^2 \left\{\textrm{sech}^2\left[\sqrt{2 \alpha} \beta
\left(t-\bar\chi-\frac{\bar\xi}{2}\right)\right] + \textrm{sech}^2\left[\sqrt{2 \alpha} \beta \left(t-\bar\chi+\frac{\bar\xi}{2}\right)\right]\right\},
\label{gxi}
\ee
%and the orthogonality condition is the same for all the points with  $\xi=0$ and read
%\be
%0&=& \left. \int~\mbox{d}t~y(t)~ g_{\chi}(t,\chi,\xi)~\right|_{{ \xi=0}}  =0\label{divT}\\
%0&=&\left. \int~\mbox{d}t~y(t)~g_{\xi}(t, \chi,\xi)~\right|_{{ \xi= 0}}  =\sqrt{2\alpha}\beta^2\int~dt~\tilde{y}(t)\textrm{sech}^{-2}\!\left[\sqrt{2\alpha}\beta \,t\right],\label{divD}
%\ee
%where we have used the fact that
%\be
%g_\xi(t; \chi, \xi)~\bigg|_{|_{ \xi=0}}  = \frac{d}{dt} \hat{x}_I(t-\chi).
%\label{gxi}
%\ee
%The fact that the condition (\ref{divT}) holds for any value $\chi$ is  a  consequence of the time translational invariance of the vacuum.  
Hence, without loss of generality, in the following we shall consider% et $\chi=0$ and develop the projection onto the one-dimensional manifold 
\be
 x(t;\chi, \xi) &=&-\beta\left\{1-\tanh\!\left[\sqrt{2\alpha}\,\beta\,\left(t-\chi+\frac{\xi}{2}\right)\right]+\tanh\!\left[\sqrt{2\alpha}\,\beta\,\left(t-\chi-\frac{\xi}{2}\right)\right]\right\}+y(t)
\label{deco}
\ee
with the conditions
\be
&&\bigg( y(t) \cdot g_\chi(t;\bar \chi,\bar\xi)\bigg) = 0,\label{ort5a}\\
&&\bigg( y(t) \cdot g_\xi(t;\bar \chi,\bar\xi) \bigg)= 0.\label{ort5b}
\ee

Although our ultimate goal is to evaluate $F_2^{IA}(\xi)$ and $F_2^{IA}(\xi)$ in a fully non-perturbative way,  it is instructive to discuss first  a  perturbative analysis, which  yields information
about the contribution to the  quantum effective interactions in the short instanton-antiinstanton distance limit. 
Such a calculation isl be presented in the next section, while the fully non-perturbative calculation is reported in section \ref{nonpert}.

\section{Perturbative Calculation}
\label{pert}

Perturbation theory  deals with  small quantum  fluctuations around a {\it classical} vacuum. In particular, a calculation
of $F_2^{IA}(\xi)$ and $F_2^{IA}(\xi)$  at small $\xi$ requires to assign to each point  in the vicinity of the trivial vacuum
\be
\tilde{x}_{S_2}\equiv-\beta
\label{triv}
\ee
 a point on the intanton-antiinstanton functional manifold. Since quantum  fluctuations  can be arbitrarily small, 
   the orthogonality conditions (\ref{ort5a}) and (\ref{ort5b}) have to be imposed at a point which is arbitrarily close to the same classical vacuum.
In principle, the most natural choice would be impose the orthogonality conditions  {\it at}   the classical vacuum. 
However, problems arise due to the fact that it is not possible to define the tangent space in such a point, since
\be
g_{\chi} (t; \bar\chi,0)\equiv0,
\ee
for all $ \bar \chi$.
To overcome this difficulty, in the following we use the stochastic quantization formalism to construct a rigorous approach in which the tangent space which is defined at a point which is 
{\it arbitrarily} close to classical point, but does not coincide with it. 

Let us begin by  briefly reviewing Pairsi and Wu quantization technique~\cite{parisi}. The starting point  is to allow the field configuration $x(t)$ to depend on an additional parameter, the so-called stochastic ''time'' $\tau$. 
The dynamics of the field in such an additional dimension is postulated to obey a Langevin equation:
\be\label{lang}
x'(t,\tau) \equiv \frac{d}{d \tau}~x(t,\tau)=-k\frac{\delta\,S[x]}{\delta\,x(t,\tau)}+\sqrt{\hbar}\,\eta(t,\tau),
\ee
where $k$ is an arbitrary diffusion coefficient and $\eta(t,\tau)$  Gaussian distributed stochastic field
\be
P[\eta]&\propto&
%\frac{1}{\int\mathcal{D}x\exp\{-\frac{1}{4k}\int_{-\infty}^\infty\mbox{d}t\int_0^\infty\mbox{d}\tau\eta^2(t,\tau)\}}
\exp\left\{-\frac{1}{4k}\int_{-\infty}^\infty\mbox{d}t\int_0^\infty\mbox{d}\tau\,\,\eta^2(t,\tau)\right\}
\ee
which obeys the fluctuation-dissipation relationship
\be
\la\eta(t, \tau)\eta(t',\tau')\ra&=&\int\mathcal{D}\eta\,\eta(t,\tau)\eta(t',\tau')P[\eta] = 2~k~\delta(t'-t)~\delta(\tau'-\tau).
\label{fluctdiss}
\ee

For any value of the stochastic time $\tau$, the probability to for the field to assume  a given configuration $x(t,\tau)$ is described by a (functional) probability distribution $\mathcal{P}[x](\tau)$, which is a solution of the Fokker-Planck Eq. associated to the Langevin Eq. (\ref{lang}):
\be
\frac{d}{ d\tau} \mathcal{P}[x] = k~\frac{\delta^2}{\delta x^2} P[x] + k~\frac{\delta}{\delta x}\left( 
\mathcal{P}[x]~ \frac{\delta S[x]}{\delta x}\right)
\label{FPE}
\ee

A general property of the Fokker Planck Eq. is that  its  solutions  converge to the static, "Boltzmann" weight, in the long time limit:
\be
\label{weight}
\mathcal{P}[x] \stackrel{(\tau\to \infty)}{\rightarrow}\frac{1}{\int\mathcal{D}x(t)
\exp\Big\{-\frac{1}{\hbar} S[x(t)]\Big\}}\exp\Big\{-\frac{1}{\hbar} S[x(t)]\Big\},
\label{Peq}
\ee
regardless of the initial condition, $x(t, \tau=0)$ and of the value of the diffusion coefficient $k$. 
Hence, the Langevin Eq. (\ref{lang}) generates configurations which, at equilibrium, 
are distributed  according to the statistical weight appearing in the Euclidean quantum path integral. 
Such configurations can be used to compute quantum mechanical Green's functions.

%\vspace{0.5 cm}
%\emph{Step 1: Generation of a Set of Perturbative Equilibrium Field Configurations}
%\vspace{0.5 cm}

In stochastic perturbation theory, a generic path $x(t,\tau)$ obeying Langevin Eq. (\ref{lang}) with boundary conditions (\ref{BC1}) is written as a power series in $\varepsilon=\sqrt{\hbar}$:
\be
%\xi(\tau)&=&\sum_{i=0}^\infty\varepsilon^i~\xi_i(\tau\label{dexpand})\\
x(t,\tau)&=&\sum_{i=i}^\infty\varepsilon^ix_i(t,\tau), 
%y(t,\tau)&=&\sum_{i=i}^\infty\varepsilon^iy_i(t,\tau)
\label{xexpand}
\ee
$x_0(t,\tau)$ is the classical content of the path, while all other terms represent quantum corrections. 
In the double-well problem, the classical solution with boundary conditions (\ref{BC3}) is
$x_0(t, \tau)= - \beta$. 

By inserting the expansion (\ref{xexpand}) into the Langevin Eq. (\ref{lang}) and matching the Left-Hand-Side (LHS) and Right-Hand-Side (RHS),  order  by order in $\varepsilon$,  one generates a tower of coupled  stochastic differential Eq.s, for the components $x_i(t, \tau)$, which appear in Eq. (\ref{xexpand}):

\be
\label{o0}
O(\varepsilon^0):&& x'_0(t, \tau)= k\,m \left(\partial_t^2 - 4\alpha ~ [x_0^2(t, \tau)-\beta^2] ~\right)~x_0(t, \tau)\\
\label{o1}
O(\varepsilon^1):&& x_1^{'}(t,\tau)=k\,m \left(\partial_t^2 - 4\alpha ~ [3\,x_0^2(t, \tau)-\beta^2] ~\right)~x_1(t,\tau)+\eta(t,\tau)\\
\label{o2}
O(\varepsilon^2):&& x_2^{'}(t,\tau)=k\,m \left[\left(\partial_t^2 - 4\alpha ~ [3\,x_0^2(t, \tau)-\beta^2] ~\right)~x_2(t,\tau)-12\alpha\, x_0(t, \tau) x_1^2(t,\tau)\right]\\
\label{o3}
O(\varepsilon^3):&& x_3^{'}(t,\tau)=k\,m \left[\left(\partial_t^2 - 4\alpha ~ [3\,x_0^2(t, \tau)-\beta^2] ~\right)~x_3(t,\tau)-4\alpha\, x_1^3(t,\tau)-24\alpha x_0(t, \tau) x_1(t,\tau)x_2(t,\tau)\right]\\
&\ldots&\nonumber
\ee

%If these Eq.s are integrated numerically, one  ensemble of field configurations $x(t, \tau)$ which is distributed according to the solution of (\ref{FPE}) at each value of the stochastic time, $\tau$.
In practice,  the perturbative expansion is truncated   %
and one solves a finite set of stochastic differential Eq.s,  
starting from a given initial condition. For example,  truncating the expansion to order $\varepsilon^2$ and choosing  the initial condition
\be
x_0(t,\tau=0)&=&-\beta,\\
x_i(t,\tau=0) &=& 0, \qquad (i=1,2,\ldots),  
\ee
which corresponds to the classical vacuum state, we find
\be
x_0(t,\tau) &=& - \beta\label{solpx0}\\
x_1(t,\tau)&=&\int_{-\infty}^{\infty}\frac{\mbox{d}\omega}{2\pi} e^{-i\omega t}\int_0^\infty\mbox{d}\tau^{'}\theta\left[\tau-\tau^{'}\right]e^{-km(8\alpha\beta^2+\omega^2)(\tau-\tau^{'})}\tilde{\eta}(\omega,\tau^{'})\label{solpx1}\\
x_2(t,\tau)&=&12\alpha\beta km\int_{-\infty}^{\infty}\frac{\mbox{d}\omega}{2\pi} e^{-i\omega t}\int_0^\infty\mbox{d}\tau^{'}\theta\left[\tau-\tau^{'}\right]e^{-km(8\alpha\beta^2+\omega^2)(\tau-\tau^{'})}~ 
\int_{-\infty}^\infty\mbox{d}t^{'}e^{i\omega t^{'}}x^{2}_1(t^{'},\tau^{'}).\label{solpx2}
\ee
The corresponding perturbative solution is 
\be
x(t,\tau)= x_0(t,\tau) + \varepsilon x_1(t,\tau) + \varepsilon^2 x_2(t,\tau).
\label{psol}
\ee
 It is important to stress that only the asymptotic equilibrium solution $x(t,\tau=\infty)$ enters in the evaluation of physical observables. Such equilibrium solutions are independent on the choice of the initial condition
of the perturbative stochastic equations (\ref{o0})-(\ref{o3}).
%To see this explicitly, let us consider e.g. Eq.(\ref{o0}). For $\tau\rightarrow\infty$ the configuration $x_0(t,\tau)$ reaches the equilibrium, i.e.
%$x_0^{'}(t,\tau=\infty)\equiv0$, which solves the Equation of Motion (EoM)
%\be
%\left. \frac{\partial S[x_0(t,\tau)]}{\partial x_0(t,\tau)}\right|_{\tau\to\infty}=0,
%\ee
%The only possible solution $x_0(t, \tau =\infty)$ of such an Eq., which satisfies the boundary condition $x_0(\pm T/2,\infty)=-\beta$ is precisely
%\be
%x_0(t,\infty)=-\beta;\quad\forall\,x_0(t,0).\label{gen0sol}
%\ee

Let us now  show how the stochastic perturbation theory technique can be used to gain information about the   $F^{IA}_2(\xi)$ and $F^{AI}_2(\xi)$ distributions. 
% At any finite stochastic time $\tau$, one obtains $l$ independent perturbative solutions of the Langevin Eq., which are entirely  specified by their $k$ quantum components: 
%\be
%&&x^{(1)}_1(t,\tau), x^{(1)}_2(t,\tau), \ldots, x^{(1)}_k(t,\tau)\label{s1}\\
%&&\ldots\nonumber\\
%&&x^{(l)}_1(t,\tau), x^{(l)}_2(t,\tau), \ldots, x^{(l)}_k(t,\tau),\label{sl}
%\ee
%Notice that,  due to the presence of the stochastic field in Eq. (\ref{o1}), each of such $l$ solutions will be different.  A
%After a sufficiently long stochastic time $\tau$,   the function  will be distributed according to the  equilibrium distribution
%\be
%\mathcal{P}_{eq}[x] \propto \exp\left\{- \frac{1}{\hbar} S\left[x(t,\tau)\right]\right\},
%\ee
%up to corrections of order $O(\varepsilon^{3})$.
%\vspace{1.5 cm}
%\emph{Step 2: Projection onto the Manifold of Instanton Field Configurations}
%\vspace{0.5 cm}
%Let us now define an algorithm which yields  the quantum probability distribution of the parameter $\xi$, to any given perturbative order $n$. 
%In order to obtain information about $F(\xi)$, one needs to project the solution (\ref{psol})
%onto the vacuum field manifold. 
To this end, we begin by decomposing the field as in Eq. (\ref{decompose}), 
\be
x(t) \equiv \tilde x_{S_2}(t;t_1,t_2) + y(t).
\ee 
Next  we need to promote the manifold field $\tilde x(t;\chi,\xi)$ and fluctuation field $y(t)$ to dynamical variables, under the stochastic time evolution.
There is some freedom associated to the definition of such a stochastic dynamics. 
For example, a possible choice may be  one in which  the $\tau$ dependence enters entirely through the fluctuation field $y(t,\tau)$, while the smooth vacuum field $\tilde x_{S_2}$ is 
assumed to be static, under stochastic evolution, i.e. $\tilde x_{S_2}(t, \tau)= \tilde x_{S_2}(t)$. 
Instead, a crucial point of the present approach is to make a different choice and allow both the fluctuation field and the smooth vacuum field to vary with the stochastic time $\tau$. 
This is done in practice by promoting  the curvilinear coordinates  $\xi$ and $\chi$ to  dynamical stochastic degrees of freedom~\cite{granati}, i.e.  $\xi\rightarrow \xi(\tau)$ and
$\chi\rightarrow \chi(\tau)$.  
Consequently, at a generic stochastic instant $\tau$, the quantum  field $x(t,\tau)$ reads: 
\be
 x(t,\tau)&=&\tilde{x}_{S_2}\left(t; \chi(\tau)-\frac{1}{2}\xi(\tau),\chi(\tau)+\frac{1}{2}\xi(\tau)\right)+y(t,\tau).
\label{key}
\ee

Let us now construct a perturbative solution  of the Langevin Eq. (\ref{lang}), based on the decomposition (\ref{key}). 
We recall that  the multi-instanton field is not a classical solution of the EoM,  except in the points where  $\xi=0$. As a consequence,   
quantum corrections will appear not only in the fluctuation field, but also in the  background field. 
To account for this fact, we  expand $y(t, \tau)$, $\chi(\tau)$ and $\xi(\tau)$ as power series in $\varepsilon=\sqrt{\hbar}$:
\be
y(t,\tau)&=&\sum_{i=1}^\infty\varepsilon^iy_i(t,\tau)\label{yexpand}\\
\chi(\tau)&=&\sum_{i=0}^\infty\varepsilon^i~\chi_i(\tau\label{mexpand})\\
\xi(\tau)&=&\sum_{i=0}^\infty\varepsilon^i~\xi_i(\tau\label{dexpand})
\ee
Let us now define the tangent space in a generic point $\bar \xi, \bar \chi$ of the manifold. It is possible to show that the orthogonality conditions (\ref{ort5a}) and (\ref{ort5b}) hold order-by-order in 
perturbation theory and at any stochastic time  i.e.:
\be
&&\bigg(y_i(t,\tau)\cdot g_\chi(t;\bar{\chi},\bar{\xi})\bigg)=0;
\label{p1}\\
&&\bigg(y_i(t,\tau)\cdot g_\xi(t;\bar\chi,\bar\xi)\bigg)=0, \qquad \forall i, \forall \tau.\label{p2}
\ee
%Such a requirement is sufficient to ensure that the decomposition in Eq. (\ref{dexpand}) and  (\ref{yexpand}) is respected by the stochastic evolution.
%We emphasize the fact that the ambiguity in the decomposition () is removed by imposing the orthogonality condition in the point $\xi=0$ of the manifold of background  fields.

From Eq.s (\ref{xexpand}), (\ref{yexpand}),  (\ref{mexpand}) and (\ref{dexpand})  it is immediate to obtain an expression for each of the $x_i(t,\tau)$ components in 
Eq. (\ref{xexpand}):%, in terms of the components of the curvilinear coordinates $\big(\chi_i(\tau),\xi_i(\tau)\big)$ and of the fluctuation fields $y_i(t, \tau)$:
\be
x_i(t,\tau) \equiv ~\left. \frac{1}{i!}\frac{\partial^i}{\partial \varepsilon^i} 
~\left( \hat x_{S_2} \left(t;\sum_{n=0}^\infty \varepsilon^n \chi_n,\sum_{m=0}^\infty \varepsilon^m \xi_m\right) + \sum_{l=1}^\infty \varepsilon^l y_l(t,\tau)~\right)~\right|_{\varepsilon=0}
\ee 
For example,  the first orders are%\footnote{ One can immediately show that $\chi_0(\tau)$ does not depend on $\tau$, $\chi_0^{'}(\tau)=0$. Hence we have set $\chi_0(\tau)=\chi_0$.}
\be
x_0(t,\tau) &=& \tilde{x}_{S_2}\left(t;\chi_0-\frac{\xi_0(\tau)}{2},\chi_0+\frac{\xi_0(\tau)}{2}\right)\label{sv0}\\
x_1(t,\tau) &=& \chi_1(\tau)~g_\chi\big(t;\chi_0,\xi_0(\tau)\big)+\xi_1(\tau)~g_\xi\big(t;\chi_0,\xi_0(\tau)\big)+y_1(t,\tau)\phantom{\bigg\{}\label{l1}\\
x_2(t,\tau) &=& \chi_2(\tau)~g_\chi\big(t;\chi_0,\xi_0(\tau)\big)+\xi_2(\tau)~g_\xi\big(t;\chi_0,\xi_0(\tau)\big)+\phantom{\bigg\{}\nonumber\\
&& -\frac{1}{2}\left(\chi_1^2(\tau)+\frac{\xi^2_1(\tau)}{4}\right)~\dot{g}_\chi\big(t,\chi_0,\xi_0(\tau)\big)-\chi_1(\tau)~\xi_1(\tau)~\dot{g}_\xi\big(t,\chi_0,\xi_0(\tau)\big)+y_2(t,\tau)\phantom{\bigg\{} \label{sv2}\\
&\ldots&.\nonumber,
\label{expand}
\ee
where we have used the fact that  $\chi_0$ is independent on $\tau$.
The terms on the LHS of Eq.s (\ref{sv0})-(\ref{sv2}) coincide with the perturbative solution results (\ref{solpx0})-(\ref{solpx2}). On the other hand,  the terms on the RHS represent the
 decomposition of the same functions in terms of the low-energy  vacuum field configurations  and of the corresponding fluctuation fields. 

In order to make contact with the effective interaction, we need to introduce the tangent space which enters the projection Eq.s (\ref{p1}) and (\ref{p2}).
At this point, we need face the above mentioned problem that the tangent space at the classical vacuum $-\beta$ is not defined. 
To overcome this problem, we let the tangent space vary with the stochastic time in such a way
that the point $\bar \xi, \bar \chi$ asymptotically approaches the classical vacuum, but does not coincide with it at any finite $\tau$.  
In practice, we promote
$\bar \xi$ to a stochastic variable and we impose
\be
\bar \xi(\tau)\stackrel{\tau\rightarrow \infty}{\rightarrow}0.
\ee  
In particular, we choose $\bar \xi(\tau) \equiv \xi_0(\tau)$, since $\xi_0(\tau\to\infty)\to0$.

 Using such a decomposition, we are now in a condition to analytically compute arbitrary moments of the  equilibrium distribution for $\chi$ and $\xi$, i.e. $\langle \xi^k\rangle$ and $\langle\chi^k\rangle$.
%We perform the inner product between the components $x_0,x_1,\ldots$ of the perturbative solution and the basis vectors of the tangent space.
%Let us discuss here the evaluation of the lowest two moments $\langle \xi\rangle$, $\langle \xi^2\rangle$.
%where we have used the property that $\frac{d}{d \xi} \hat{x}_I(t;\xi) = - \frac{d}{d t} \hat{x}_I(t;\xi)$.
%By looking at Eq.s (\ref{expand}) we can conclude that it is a natural choice to set $(\bar\chi=\chi_0,\bar\xi=\xi_0(\tau))$, that is to say project in the tangent space built on the coordinates of the ''classical part'' of the configuration. In fact, for every choice of $\chi$ and $\xi$ we have the following relations
%\be
%&&\bigg(g_\chi(t;\chi,\xi)\cdot g_\xi(t;\chi,\xi)\bigg)\equiv0;\\
%&&\bigg(\dot{g}_\chi(t;\chi,\xi)\cdot  g_\chi(t;\chi,\xi)\bigg)\equiv0;\\
%&&\bigg(\dot{g}_\xi(t;\chi,\xi)\cdot g_\xi(t;\chi,\xi)\bigg)\equiv0.
%ee
%Note that wIn particular, if we expand around the classical vacuum $\xi_0(\tau)=0$, then these 
By projecting and inverting Eq.s (\ref{l1}) and (\ref{sv2}), we obtain the following expression for the collectives coordinates up to $\mathcal{O}(\hbar)$ %Eq.s (\ref{expand}) simplify to
%\be
%x_0(t,\tau)&=& -\beta\label{sv0}\\
%x_1(t,\tau)&=& \xi_1(\tau)g_\xi(t,0)+y_1(t,\tau)\label{sv1}\\
%x_2(t,\tau)&=& \xi_2(\tau)g_\xi(t,0)+y_2(t,\tau) \label{sv2}\\
%x_3(t,\tau)&=& \xi_3(\tau)g_\xi(t,0)+\frac{1}{24}\xi_1^3(\tau) \ddot g_\xi%(t,0)+y_3(t,\tau)\label{sv3}\\
%&\ldots&\nonumber
%\label{expand2}
%\ee
%The terms on the LHS of Eq.s (\ref{sv0})-(\ref{sv2}) represent the perturbative solution results (\ref{solpx0})-(\ref{solpx2}), while  the terms on the RHS represent their decomposition in terms of the low-energy  vacuum field configurations  and of the fluctuation fields. 
% in the origin $g_\xi(t, 0)$,
% and using the set of orthogonality conditions (\ref{ort3}), one finds:
\be
%\xi_0(\tau) &=& 0\label{xi0}\\
\chi_1(\tau)&=&\frac{\left(x_1(t,\tau)\cdot g_\chi(t;\chi_0,\xi_0(\tau))\right)}{\left(g_\chi(t;\chi_0,\xi_0(\tau)\cdot g_\chi(t;\chi_0,\xi_0(\tau)\right)}\label{xi1},\\
\xi_1(\tau)&=&\frac{\left(x_1(t,\tau)\cdot g_\xi(t;\chi_0,\xi_0(\tau))\right)}{\left(g_\xi(t;\chi_0,\xi_0(\tau)\cdot g_\xi(t;\chi_0,\xi_0(\tau)\right)}\label{xi2},\\
\chi_2(\tau)&=&\frac{\left(x_2(t,\tau)\cdot g_\chi(t;\chi_0,\xi_0(\tau))\right)+\chi_1(\tau)\xi_1(\tau)\left(\dot{g}_\xi(t;\chi_0,\xi_0(\tau))\cdot g_\chi(t;\chi_0,\xi_0(\tau))\right)}{\left(g_\chi(t;\chi_0,\xi_0(\tau))\cdot g_\chi(t;\chi_0,\xi_0(\tau))\right)},\\
\xi_2(\tau)&=&\frac{\left(x_2(t,\tau)\cdot g_\xi(t;\chi_0,\xi_0(\tau))\right)+\frac{1}{2}\left(\chi_1^2(\tau)+\frac{1}{2}\xi^2_1(\tau)\right)\left(\dot{g}_\chi(t;\chi_0,\xi_0(\tau))\cdot g_\xi(t;\chi_0,\xi_0(\tau))\right)}{\left(g_\xi(t;\chi_0,\xi_0(\tau)\cdot g_\xi(t;\chi_0,\xi_0(\tau)\right)},\\
%\xi_3(\tau)&=&\frac{3}{4}\frac{\left(x_2(t,\tau),g_\xi(t,0)\right)}{\sqrt{2\alpha}\beta^3}-\frac{16}{135}\xi_1^3(\tau)\alpha\beta^2\label{xi3}\nonumber\\
\ldots &&.
\ee
%where we have used the following integrals
%\be
%\left(g_\xi(t,0),g_\xi(t,0)\right)&=&\frac{4}{3}\sqrt{2\alpha}\beta^3\\
%\left(\dot g_\xi(t,0), g_\xi(t,0)\right) &=& 0.
%\left(\ddot g_\xi(t,0),g_\xi(t,0)\right)&=&-\frac{32}{15}\sqrt{2\alpha}\alpha\beta^5.
%\ee
Using the fluctuation-dissipation relationships (\ref{fluctdiss}), and the fact that $\xi_0(\tau)$ is independent from $\eta(t,\tau)$ we find 
\be
\langle\chi_1(\tau)\rangle&=&0\\
\langle\chi_2(\tau)\rangle&=&0\\
\langle\xi_1(\tau)\rangle&=&0\\
\langle\xi_2(\tau)\rangle&=&\frac{\langle x_2(t,\tau)\rangle\left(1\cdot g_\xi(t;\chi_0,\xi_0(\tau))\right)+\frac{1}{2}\left(\langle\chi_1^2(\tau)\rangle+\frac{1}{2}\langle\xi^2_1(\tau)\rangle\right)\left(\dot{g}_\chi(t;\chi_0,\xi_0(\tau))\cdot g_\xi(t;\chi_0,\xi_0(\tau))\right)}{\left(g_\xi(t;\chi_0,\xi_0(\tau)\cdot g_\xi(t;\chi_0,\xi_0(\tau)\right)}\nonumber\\
&\bigg\downarrow&\tau\rightarrow\infty\nonumber\\
&=&\frac{9}{32}\frac{1}{m\alpha\beta^4}
\ee
Hence, we have obtained a closed analytical expression for the first moments:
\be
\langle\chi\rangle &=& \langle\chi_0 + \varepsilon \chi_1 + \varepsilon^2 \chi_2 \rangle = \chi_0 + O(\hbar^2)\\
\langle\xi\rangle &=& \langle\xi_0 + \varepsilon \xi_1 + \varepsilon^2 \xi_2 \rangle = \frac{9}{32}\frac{\hbar}{m\alpha\beta^4} + O(\hbar^2)
\ee
Now, in order to compute the second moments, we observe that the general expression up to order $O(\varepsilon^2)$ is
\be
\chi^2(\tau)&=&\chi^2_0(\tau)+2\varepsilon\chi_0\chi_1(\tau)+\varepsilon^2[2\chi_0\chi_2(\tau)+\chi_1^2(\tau)]+\ldots\\
\xi^2(\tau)&=&\xi^2_0(\tau)+2\varepsilon\xi_0(\tau)\xi_1(\tau)+\varepsilon^2[2\xi_0(\tau)\xi_2(\tau)+\xi_1^2(\tau)]+\ldots
\ee
which immediately gives 
\be
\langle\chi^2\rangle&=&\infty;\label{q1}\\
\langle\xi^2\rangle&=&\hbar\langle\xi_1^2\rangle=\frac{9}{32}\frac{\hbar}{m\alpha\beta^4}\left(\frac{\pi^2-9}{3\sqrt{2\alpha}\beta}\right).
\label{q2}
\ee

Some comments on these results are in order. 
The distribution of the instanton-antiinstanton distance $\xi$ is not symmetric around the origin,  since $\langle \xi \rangle \ne 0$.   
Such a symmetry breaking  comes from  fluctuations which explore the non-harmonic region of the potential function $U(x)$. Since the potential on the left of the equilibrium configuration raises more steeply 
than that on the  right, 
quantum  paths in the direction of the barrier are statistically favored. 
%Consequently, once such perturbative quantum fluctuations are projected onto the low-dimensional functional manifold spanned by instanton-antiinstanton configurations, the effect of the breaking of reflection symmetry around $-\beta$  is to enhance the probability of fluctons with $t_1<t_2$, that is to favor fluctons made by a close instanton-antiinstanton pair,
%over  fluctuons made by a close antiinstanton-instanton pair. 
The divergence  $\langle \chi^2\rangle=\infty$ emerges because the distribution of collective coordinates is independent on $\chi$,
 as consequence of the time-translational invariance of the system. In the language of stochastic  quantization, this implies that
the center of mass of the instanton-antiinstanton pair performs Brownian motion in stochastic time and $\langle \chi^{2}\rangle\propto \tau$, according to Einstein 
relationship. 

We emphasize once again that in this calculation we have never requested that the multi-instanton configurations should be approximate solutions of the classical EoM.
The only request is that the configuration corresponding to the classical vacuum must belong to the manifold parametrized by the relevant low-energy degrees of freedom. 
We also stress the fact that there is no contribution to the effective interaction, at the classical level: $F_2(\xi)$ is an entirely  quantum effect.   

\section{Non-Perturbative Calculation}
\label{nonpert}

Let us now take the main step of the present work and  perform a fully non-perturbative calculation of $F_2(\xi)$ which describes the correlations between 
consecutive tunneling events.

%extract a specific value for the parameters $\xi_i$, which enter in the perturbative expansion of $\xi$, Eq. (\ref{dexpand}) 
%from each of the $l$ independent solutions of the set of coupled Eq.s (\ref{o0})-(\ref{o3}).
%When this procedure is repeated for a large number $l$ of independent  equilibrium configurations, one determines a perturbative estimate of the quantum probability distribution of 
%the instanton-antiinstanton distance $\xi$:
%\be
%\mathcal{P}_{eq}[\xi] = \lim_{\tau\to\infty}~\mathcal{P}\left[\sum_{i=1}^n\varepsilon^i\xi_i(\tau\to \infty)\right]+O(\varepsilon^{n+1})
%\ee
%The perturbative expression of the effective  action for $\xi$ in the vicinity of the classical configuration $\xi=0$ is obtained by taking the logarithm of such a probability distribution:

Let  $\{x^{1}(t,\tau=\infty), \ldots, x^{l}(t,\tau=\infty)\}$  be an ensemble of $l$  equilibrium field configurations, which were obtained  non-perturbatively, for example by integrating numerically directly the Langevin Eq. (\ref{lang}), or by means of a lattice Monte Carlo simulation.% These are just the ordinary configurations used in all lattice field theory calculations.
The pair correlation function $g_2^{IA}(\xi)$ can be extracted by projecting the set of equilibrium configurations onto the  low-energy vacuum field manifold spanned by an instanton-antiinstanton pair.  
To this end, we define the functionals of the field configuration $x(t,\tau)$
\be
\Phi_\chi[x_\tau]&:=&\left(x(t,\tau),g_\chi(t;\bar\chi,\bar\xi)\right),\label{e1}\\
\Phi_\xi[x_\tau]&:=&\left(x(t,\tau),g_\xi(t;\bar\chi,\bar\xi)\right),\label{e2}
\ee
which represents the projection of an arbitrary field configuration onto the tangent space, at the point ($\bar\chi,\bar\xi$). 
We also introduce
the functions of the collective coordinate $\chi$ and $\xi$
\be
\Psi_\chi(\chi,\xi) &:=&\left(\tilde{x}_{S2}\left(t,\chi-\frac{1}{2}\xi,\chi+\frac{1}{2}\xi\right),g_\chi(t;\bar\chi,\bar\xi)\right),\\
\Psi_\xi(\chi,\xi) &:=&\left(\tilde{x}_{S2}\left(t,\chi-\frac{1}{2}\xi,\chi+\frac{1}{2}\xi\right),g_\xi(t;\bar\chi,\bar\xi)\right),
\ee
which represents the projection of a generic point of the instanton-antiinstanton field manifold onto the same tangent space. In the specific  case of the double-well potential one has
\be
\Psi_\chi(\chi,\xi)&=&\zeta\left[\left(\chi-\frac{\xi}{2}\right)-\left(\bar\chi+\frac{\bar\xi}{2}\right)\right]-\zeta\left[\left(\chi+\frac{\xi}{2}\right)-\left(\bar\chi+\frac{\bar\xi}{2}\right)\right]+\nonumber\\
&&\!+\,\zeta\left[\left(\chi+\frac{\xi}{2}\right)-\left(\bar\chi-\frac{\bar\xi}{2}\right)\right]-\zeta\left[\left(\chi-\frac{\xi}{2}\right)-\left(\bar\chi-\frac{\bar\xi}{2}\right)\right],\\
\Psi_\xi(\chi,\xi)&=&-2\beta^2+\frac{1}{2}\zeta\left[\left(\chi-\frac{\xi}{2}\right)-\left(\bar\chi+\frac{\bar\xi}{2}\right)\right]-\frac{1}{2}\zeta\left[\left(\chi+\frac{\xi}{2}\right)-\left(\bar\chi+\frac{\bar\xi}{2}\right)\right]+\nonumber\\
&&\!-\,\frac{1}{2}\zeta\left[\left(\chi-\frac{\xi}{2}\right)-\left(\bar\chi+\frac{\bar\xi}{2}\right)\right]+\frac{1}{2}\zeta\left[\left(\chi+\frac{\xi}{2}\right)-\left(\bar\chi+\frac{\bar\xi}{2}\right)\right],
\ee
where
\be
\zeta[X]=2\beta^2\left\{\sqrt{2\alpha}\beta X\sinh^{-2}\left[\sqrt{2\alpha}\beta X\right]-\coth\left[\sqrt{2\alpha}\beta X\right]\right\}.
\ee

By setting Eq.s (\ref{e1}) and (\ref{e2}) to be equal to $\Psi_\chi$ and $\Psi_\xi$ respectively, we obtain a complete system of equations for the variables $\chi$ and $\xi$.
\be
\label{system}
\left\{
\begin{matrix}
\Phi_\chi[x_\tau]&\equiv&\Psi_\chi\big[\chi(\tau),\xi(\tau)\big]\phantom{\bigg\{}\\
\Phi_\xi[x_\tau]&\equiv&\Psi_\xi\big[\chi(\tau),\xi(\tau)\big].\phantom{\bigg\{}
\end{matrix}
\right.
\ee

Such a  system has  a unique solution for  any choice of the projection point $(\bar\chi,\bar\xi)$, with $\bar\xi\ne0$.
% In fact, as we have seen in section \ref{doublewell}, for this particular choice $\Psi_\chi\equiv0$ and the system has no solution.
Hence,  it is possible to  assign a value of $\chi$ and $\xi$ to every non-perturbatively generated configuration $x(t,\tau)$.
%\be
%\xi(\tau)=\Psi^{-1}\left[\Phi[x_\tau]\right].
%\ee

Repeating such a projection for the entire ensemble of equilibrium configurations,  one obtains an histogram which by construction  is proportional to the pair correlation function 
$g^{IA}_2(\xi)$.  The effective potential $F_2(\xi)$ is immediately extracted from:
\be
\label{Peqxi}
F^{IA}_2(\xi) &=& -\hbar~\log[~g^{IA}_2(\xi)] + \textrm{const.}.
\ee
Clearly, the calculation of $F^{AI}_2(\xi)$ would be completely analog. In practice, such a calculation is not necessary, since the function $F^{AI}_2(\xi)$ can be inferred directly by symmetry arguments:
\be
F^{AI}_2(\xi) = F^{IA}_2(-\xi).
\ee
Once the effective interaction has been determined, one can evaluate the   instanton density of the liquid by minimizing the free-energy of the ensemble.
In the next section, we present the results of some numerical investigations, in order to illustrate the method and assess the accuracy of the determination of the instanton-antiiinstanton interaction.

\section{Numerical Studies}
\label{test}
\begin{figure}[t]
\includegraphics[width=9cm]{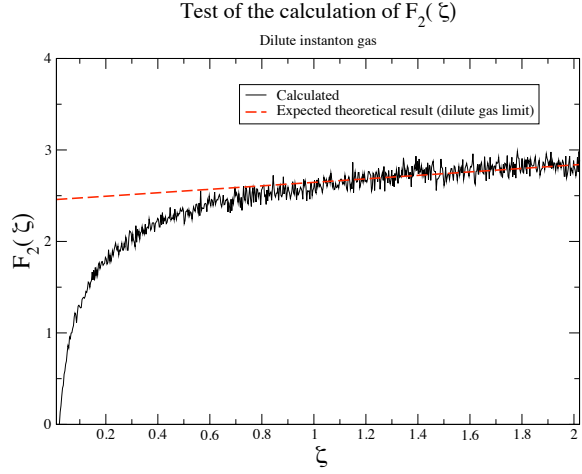}
\caption{Non-perturbative calculation of the effective interaction $F_2(\xi)$ for a dilute instanton gas with $\alpha=7$, $m=1$, $\beta=1$ in a volume $T=200$. The points are the results 
obtained from projecting 1000 configurations, while the dashed line is the expected theoretical results for $\xi$ much larger than the instanton size (which is 0.26, in these units). }
\label{checkDIGA}
\end{figure}

In order to show that our method yields the correct result, let us first  consider a  model for which the effective interaction can be evaluated analytically.
The two-body part of the effective interaction for a dilute instanton gas can 
be easily computed from Eq. (\ref{FDIGA}), by integrating out all the collective coordinates, except for those of a single instanton-antiinstanton pair.  The result is
\be
e^{-\frac 1\hbar F_2^{IA}(\xi)} =  \textrm{const.} \times e^{- (\kappa T-1) \log\left(T-\xi \right)},
\label{F2DIGA}
\ee
This result holds  for high barriers and distances $\xi$ much larger than  the instanton size. Notice that, in the thermodynamic limit --- $N, T\to \infty$ and $N/T=\kappa$ fixed--- the effective interaction $F_{2}^{IA}(\xi)$ should scale linearly, with a slope controlled by the instanton rate $\kappa$.

We now address the question if our projection technique is able to reconstruct  the effective interaction in Eq.(\ref{F2DIGA}). To this end,
 we have generated an ensemble of 1000 dilute gas configurations, by randomly sampling the positions of instantons and  anti-instantons, in a box of size $T=200$ for a well with $\alpha=7, m=1, \beta=1$.
In Fig.~\ref{checkDIGA}, we compare the expected theoretical curve (dashed line)  with the result of our numerical calculation (points). We see that, as soon as the distance $\xi$ becomes larger than few instanton sizes ---which is 0.26 in this units--- the numerical results agree with the expected curve. A linear fit of the data for $\xi> 1$ yields 
a slope of $0.32\pm 1$, in excellent agreement with the exact theoretical result, which is $0.31$. 
Hence, we conclude that our projection method is indeed able to quantitively reconstruct the structure of the exact distribution used to generate the ensemble of configurations.

Let us now discuss for completeness the structure of the effective interaction,  for our original  quantum double-well system. At this level, we no longer consider the semi-classical dilute gas model. Instead,  we account for quantum 
fluctuations to all orders.  As the barrier becomes higher and higher, performing a sampling of multiple barrier-crossing paths contributing to the functional integral with dynamical algorithms such as Molecular Dynamics of Monte Carlo
becomes highly inefficient\footnote{This difficulty is not present in QCD, where it has been shown that typical lattice configurations contain indeed many instantons and anti-instantons.}, and computationally expensive. 
To cope with this problem, we have evaluated the instanton-antiinstanton interactions using 
 the importance sampling approach described in appendix~\ref{algorithm}.     
 
Fig. \ref{npFig} shows  the results of such a non-perturbative calculation for a well with  $\alpha=1$ (low barrier) and $\alpha= 7$ (high barrier).
Some comments on these results are in order. First of all we note that the minimum of the effective interaction $F^{IA}_2(\xi)$ is located at positive values of $\xi$,  in qualitative agreement with our perturbative calculation. The range of  $\xi$ in which the effective interaction $F^{IA}_2(\xi)$ is not flat corresponds to close, largely overlapping instanton-antiinstanton pair configurations. When the distance becomes of the order of twice the instanton size, the effective interaction starts to raise and eventually reaches the dilute gas limit.
On the other hand, for low barriers, the instanton density is large and the 
attraction and repulsion generated by $F^{IA}_{2}(\xi)$ become important. 
In such a regime, the vacuum 
behaves like a one-dimensional  liquid,  rather than as an ideal gas. We note that this is precisely the physical picture underlying the instanton liquid model
of the  vacuum~\cite{fluct}.

\begin{figure}[t]
\includegraphics[width=7cm]{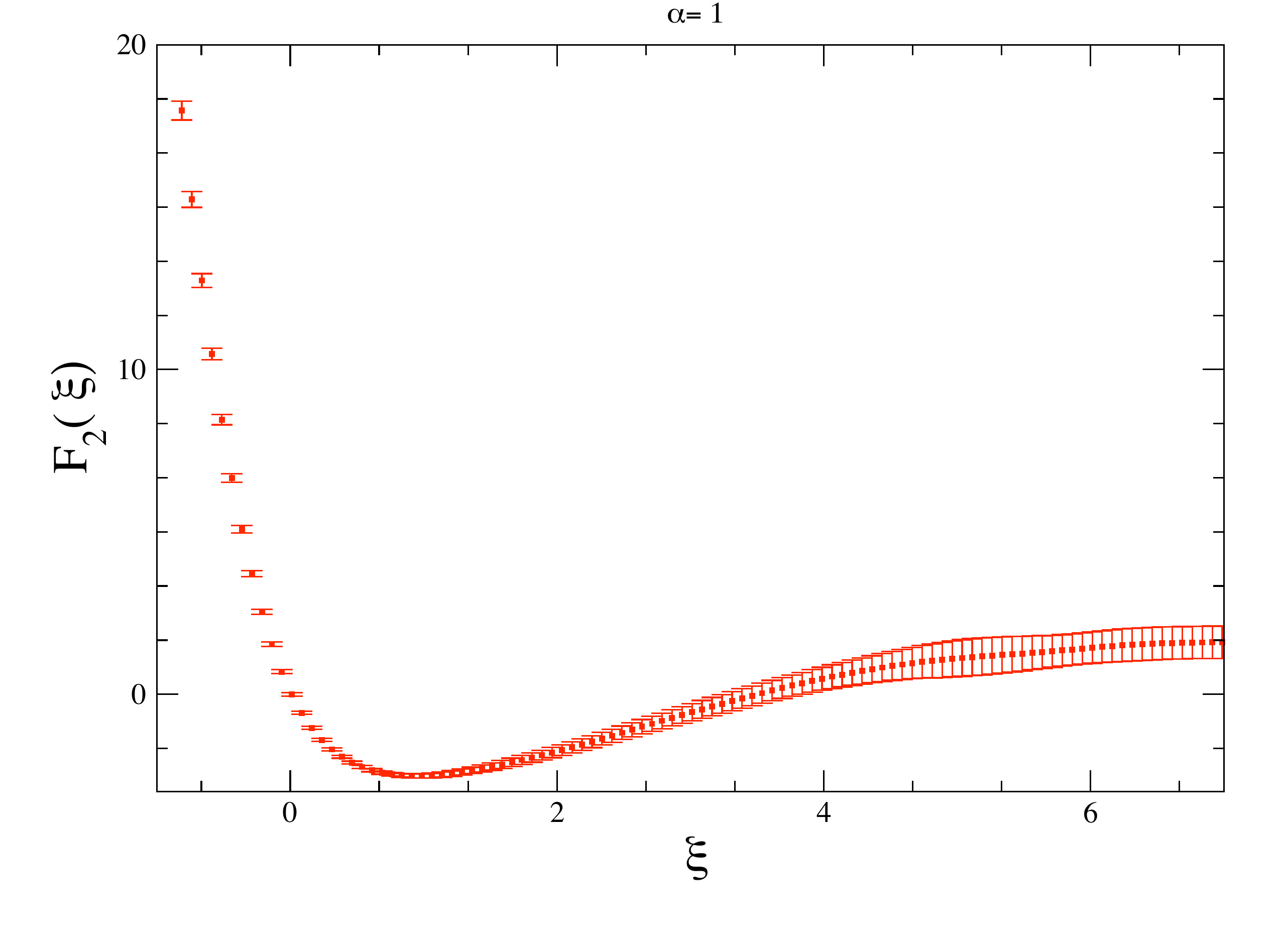}
\includegraphics[width=7cm]{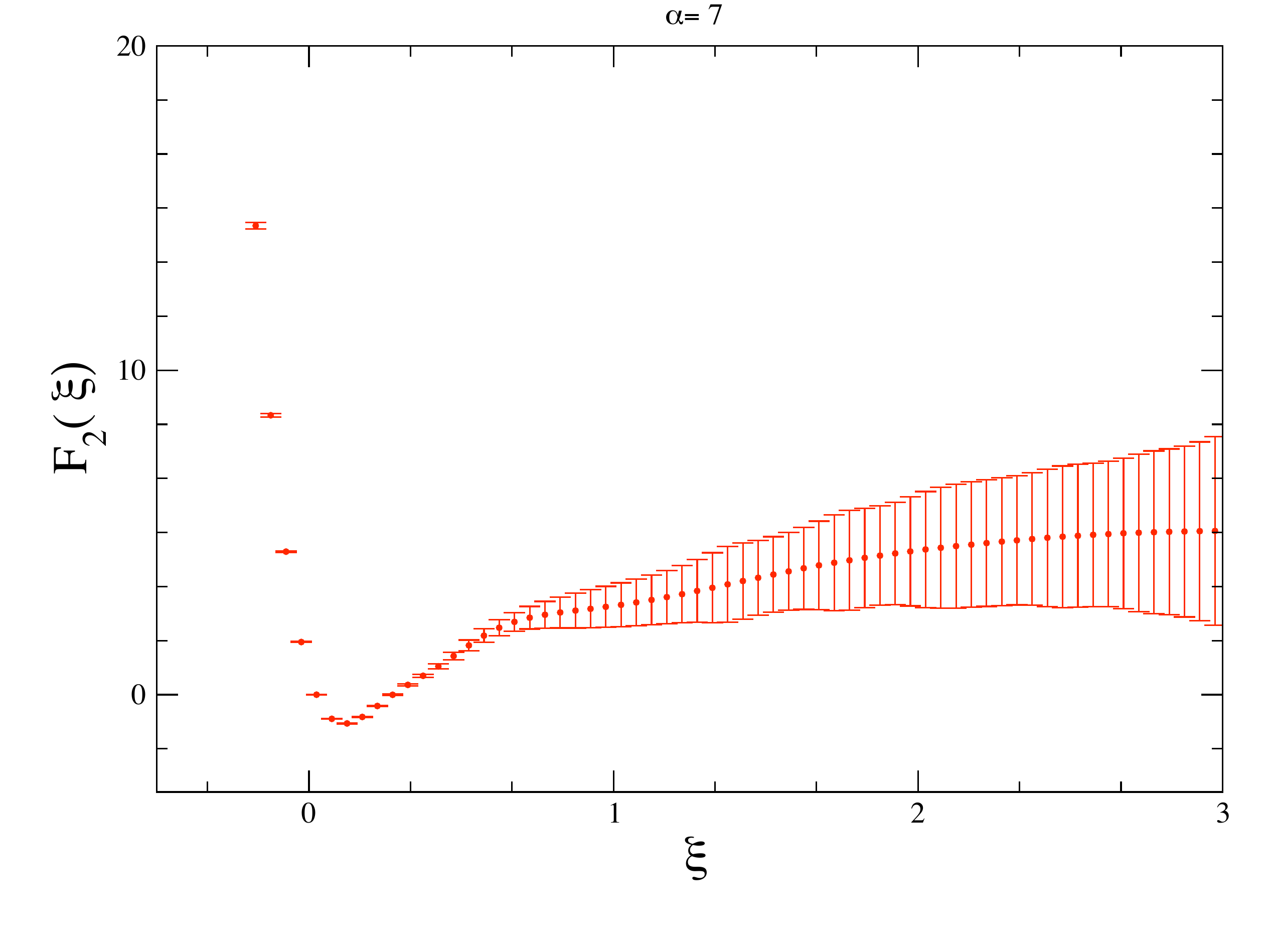} 
\caption{Non-perturbative calculation of the effective interaction $F(\xi)$ for $\alpha=1$ (left panel) and $\alpha = 7$. }
\label{npFig}
\end{figure}

\section{Conclusions}
\label{conclusions}

In this paper, we have presented an approach which allows to express quantum mechanical  path integrals,  in terms of few ordinary integrals over a set of low-energy variables, 
which parametrize the manifold of the relevant vacuum field configurations.
We have developed a rigorous technique to  extract the effective interaction, a simple quantum mechanical problem, in which the low-energy degrees of freedom are multi-instanton configurations.

We have assessed the accuracy of our method by correctly  reconstructing the effective interaction used to generate an ensemble of sinthetic configurations. We have also performed both perturbative and non-perturbative calculation of the quantum effective interaction of an instanton-antiinstanton pair.   In both cases, we have found that the effect of quantum fluctuations is to shift the location of the minimum of the effective interaction $F^{IA}_2(\xi)$ to the right. 
 
We stress the fact that, although the present discussion has focused on an instanton liquid picture of the vacuum, our projection method does not rely at all on the semi-classical approximation. The semi-classical approximation has been used only as a guidance to find good vacuum effective degrees of freedom, for our toy model. Hence, the method can in principle be generalized to build effective theories for the vacuum,  based on different types of field configurations. This observation may become important in QCD, where fields other than singular gauge  instantons are needed, in order  to account for confinement.

If the vacuum fields selected are the ones driving the system's non-perturbative dynamics, then one expects that the contribution coming to the fluctuations around them  
to the field operators appearing in the Green's functions  will be small. In this case, the calculations of observables in the effective theory can be performed very efficiently, because they involve only few ordinary integrals over the set of curvilinear coordinates. 

Most importantly, once a specific choice of the vacuum manifold has been identified, the corresponding effective theory  yields parameter-free predictions.
Hence, the present framework can be used to  assess the importance of different families of vacuum fields,  by directly comparing the results of the corresponding effective theory with the experimental data. 

The extension of the present formalism to QCD is in progress and will be presented in our upcoming papers.

\appendix

\section{Algorithm Used in the Evaluation of $F_{2}^{IA}(\xi)$}
\label{algorithm}

We are interested in computing numerically the integral
\be
g^{IA}_2(\xi) &=& \mathcal{N}~\int~\mathcal{D} y~\exp\left\{-\int~\mbox{d}t~\mathcal{L}[\tilde{x}_{S_2}(t; \chi-\frac{1}{2}\xi,\chi+\frac{1}{2}\xi)+y(t)]\right\}\delta\bigg( y\cdot g_{\chi}\big(\bar{\chi},\bar{\xi}\big)\bigg)\delta\bigg( y\cdot g_{\xi}\big(\bar{\chi},\bar{\xi}\big)\bigg)\times\nonumber\\
&&\!\!\!\!\!\!\!\!\!\!\!\!\!\!\!\!\!\!\!\!\!\!\!\!\!\!\!\!\!\!\times\bigg[\bigg( g_{\chi}\big(\chi,\xi\big)\cdot g_{\chi}\big(\bar{\chi},\bar{\xi}\big)\bigg)\bigg( g_{\xi}\big(\chi,\xi\big)\cdot g_{\xi}\big(\bar{\chi},\bar{\xi}\big)\bigg)-\bigg( g_{\chi}\big(\chi,\xi\big)\cdot g_{\xi}\big(\bar{\chi},\bar{\xi}\big)\bigg)\bigg( g_{\xi}\big(\chi,\xi\big)\cdot g_{\chi}\big(\bar{\chi},\bar{\xi}\big)\bigg)\bigg],\label{inta}
\ee
where the term inside the square brackets is the explicit representation of the Jacobian factor $\Phi[y]$.

The meta-stability of the  double well system makes it rather computationally challenging to generate a statistically 
significative ensemble of field configurations,  using algorithms based on Metropolis or by Langevin dynamics. 
The main problem is that, for such a meta-stable system,  ergodicity is reached only in an exponentially large computational time.

The problem has no easy solution within a dynamical Monte Carlo approach.  However, because of the low
dimensionality of our system, simpler importance sampling technique are available and 
efficient\footnote{Note that this problem is unrelated to our projection approach, which is a prescription
about the measurement of an effective action, once a significant sample of configurations has been provided in
some way.}.
Since the system is time-translationally invariant, without loss of generality we can set $\chi=0$, $\bar\chi=0$ --- i.e. we can remove completely the dependance from the center of mass---, and set $\bar\xi=0$. Then, the resulting expression for the pair correlation function can be re-written as
\be
g^{IA}_2(\xi)=\mathcal{N}\int\mathcal{D}y~\delta(y(t)\cdot g_\xi(t;0))~ \Phi[y]~\hat P[y(t)]~\frac{e^{-S\left[\tilde{x}_{S_2}\left(t,-\frac{1}{2}\xi,\frac{1}{2}\xi\right)+y(t)\right]}}{\hat P[y(t)]},
\label{gIA}
\ee 
where $\hat P[y(t)]$ is a probability distribution to be defined below. We stress that now the integral can be
restricted to the small region in which the projection function is not exponentially small.  The discretized
version of Eq.~(\ref{gIA}) is
\be
g^{IA}_2(\xi)= \mathcal{N}~\int \prod_{k=1}^N \mbox{d}y(t_k)\delta\left(\sum_{k=1}^Ny(t_k)g_\xi(t_k,0)\right)~ \Phi[y]~\hat P[y]~\frac{e^{-S_{\rm lat}\left[\tilde{x}_{S_2}\left(t_k,-\frac{1}{2}\xi,\frac{1}{2}\xi\right)+y\right]}}{\hat P[y]},
\label{gIAdisc}
\ee
where $N$ is the number of points in the lattice and $S_{\rm lat}$ is the discretised version of $S$. 

For $\hat P[y]$ we choose:
\be
\hat P[y]\propto \exp\left(-\frac{1}{8 m \Delta t} \sum_{k=0}^N (y(t_{k+1})-y(t_k))^2 \right)
\ee
with the constraint $y(t_0)=y(t_{N+1})=0$. 
We eliminate the delta function by setting the last coordinate $y(t_N)$ equal to
\be
y(t_N)=-\sum_{i=1}^{N-1}y(t_i)g_\xi(t_i)/g_\xi(t_N).  
\ee
Notice that, in this way, the orthogonality condition is satisfied configuration by configurations.
The statistical weight of resulting each paths was evaluated from 
\be
w_i(\xi)=\frac{\exp(-S_{\rm lat}[\tilde{x}_{s}(t_k, \xi)+y_i])}{\hat{P}[y_i]}.
\ee
Up to an overall multiplicative factor,  $g^{IA}(x)$ can be extracted from 
\be
g^{IA}_2(\xi)=\textrm{const.}\times \sum_iw_i(\xi)
\ee
By taking the logarithm, one obtains $F^{IA}_2(\xi)$, up to an overall additive constant.

%In order to deal with this difficulty, we have adopted a different strategy to evaluate $F^{IA}_{2}(\xi)$ for this system. The main point of our strategy consists in observing that 
%the function used to project onto the vacuum field manifold is exponentially suppressed for values of $t$ much larger than the typical instanton size 
%$\simeq1/(2\sqrt{2\alpha}\beta)$:
%\be
%g_\xi(t;0) \stackrel{t\rightarrow\infty}{\longrightarrow}4\sqrt{2\alpha}\beta^2 e^{-2\sqrt{2\alpha}\beta\,t}.
%\ee
%This means that the projection procedure is sensitive only to a small part $[-\tilde T, \tilde T]$ of the full $[-T,T]$ time domain. 
%Hence, the structure of the field configurations  far from such a window  do not significantly contribute to the projection function
%\be
%\Phi[x]=(x(t),g_\xi)=\left(\tilde{x}_{S_2}\left(t,-\frac{1}{2}\xi,\frac{1}{2}\xi\right),g_\xi\right)=\Psi[\xi]
%\ee
%and can be neglected. Notice that inside the projection window  $[-\tilde T, \tilde T]$ one typically expects to find at most a single pseudo-particle pair.

%Based on such observations, we have restricted the time integration to the region $[-\tilde{T},\tilde{T}]$% and considered the probability to find in such a window instanton-antiinstanton 
%pair with distance $\xi$, relative to the probability of finding a pair of fixed distance $\xi_0$,
%\be
%R(\xi)\equiv\frac{g_2^{IA}(\xi)}{g_2^{IA}(\xi_0)}.
%\label{ratio}
%\ee
%From the logarithm such a ratio of probabilities, one can extract $F_2^{IA}(\xi)$, up to an addictive constant.
%Based on such considerations,  we have re-written 

\acknowledgements
This work was motivated by inspiring discussions with F. Di Renzo.  We thank also F.Pederiva for numerical help. 
Part of this work was performed when P.F. was visiting I.Ph.T. of C.E.A. (Saclay) under a C.N.R.S. grant.  

\end{document}